\documentstyle[12pt]{article}

\newcommand{\be}{\begin{equation}}
\newcommand{\ee}{\end{equation}}
\newcommand{\bea}{\begin{eqnarray}}
\newcommand{\eea}{\end{eqnarray}}

\newcommand{\V} {{\cal V}_{n}}

\begin{document}

\begin{center}
\begin{large}
{\bf   A Holographic Interpretation   \\}
{\bf   of  \\}
{\bf  Asymptotically de Sitter Spacetimes  \\}
\end{large}  
\end{center}
\vspace*{0.50cm}
\begin{center}
{\sl by\\}
\vspace*{1.00cm}
{\bf A.J.M. Medved\\}
\vspace*{1.00cm}
{\sl
Department of Physics and Theoretical Physics Institute\\
University of Alberta\\
Edmonton, Canada T6G-2J1\\
{[e-mail: amedved@phys.ualberta.ca]}}\\
\end{center}
\bigskip\noindent
\begin{center}
\begin{large}
{\bf
ABSTRACT
}
\end{large}
\end{center}
\vspace*{0.50cm}
\par
\noindent
One of the remarkable outcomes of the AdS/CFT correspondence
has been the generalization of Cardy's entropy formula for
arbitrary dimensionality, as well as  a variety of  anti-de Sitter
scenarios. More recently, related work has
been done  in the realm of  asymptotically  de Sitter
backgrounds. Such studies presume a well-defined dS/CFT duality,
which has not yet attained the credibility of its AdS analogue.
In this paper, we  
 derive and interpret generalized forms of the Cardy entropy for a selection
of asymptotically de Sitter spacetimes. These include
  the Schwarzschild-de Sitter black hole (as a review of hep-th/0112093), 
the Reissner-Nordstrom-de
Sitter black hole and a special class of topological de Sitter
solutions.  Each of these cases is found to have interesting implications
in the context of the proposed correspondence.

\newpage

\section{Introduction}

It is commonly believed that the holographic principle
will be an essential constituent of  any  valid
theory of quantum gravity \cite{tho,sus}.  This principle
(which follows from  Bekenstein's entropy bound \cite{bek2}) 
implies that,  if  given a physical system, the relevant degrees
of freedom must suitably live  on a surface  enclosing that
system.
To take it a step further, given a physical  theory in $d$+1-dimensional
spacetime, one would expect  the existence of a $d$-dimensional theory
that can capably describe the same physics. 
\par
In spite of its esoteric origins,
the holographic principle has  explicitly been  realized 
by way of the    AdS/CFT correspondence \cite{mal,gub}.
That is,  a duality has been demonstrated between $d$+1-dimensional 
anti-de Sitter (AdS) spacetimes and appropriately defined,
$d$-dimensional conformal field theories (CFTs). 
More specifically, the thermodynamics at the horizon
of an AdS black hole can be identified with the
thermodynamics  of a CFT that lives on
an asymptotic boundary of the AdS background \cite{wit}.
\par
Recently, Verlinde \cite{ver}
 has  utilized  this AdS/CFT duality 
to demonstrate  a  result of considerable significance.
This author was able to generalize Cardy's well-known
entropy formula (for a 2-dimensional CFT) \cite{car} to
a theory of arbitrary dimensionality. The essence
of Verlinde's work was the identification of Cardy's 
central charge (the number of massless particle species)
with the Casimir (or sub-extensive) contribution to
the CFT energy. Of further significance, this Casimir energy
can  be fixed,  precisely, by way of the AdS/CFT duality 
(provided that the conformal symmetry at the boundary
 has  properly been exploited).
\par
This generalized Cardy entropy, or Cardy-Verlinde formula,
has since been extended  to  a variety of asymptotically
AdS spacetimes (see, for instance, Ref.\cite{cai}),
as well as  dynamical-boundary scenarios
(see, for instance, Ref.\cite{sav}).\footnote{There
has been an abundance of work in these areas. One
might consult  Ref.\cite{odie} for a  listing
 of most of the relevant citations.}
The range of validity is quite remarkable given
the  ``modest''  scope of the original Cardy formula.
\par
In analogy to the AdS/CFT duality, a correspondence
between de Sitter (dS) spacetimes and CFTs has
similarly been proposed \cite{str2}. (For earlier
work along this line, see Refs.\cite{first3}-\cite{last3}.)
At the simplest level, any dS spacetime is related to an AdS space by
a trivial   reversal in the sign of the cosmological constant.
However, this sign change   has significant
repercussions  that can  hinder  a quantum (or semi-classical) interpretation
of dS geometries \cite{ssv}. For instance, dS spacetimes lack
a globally timelike Killing vector, a spatial infinity,
 an objective observer and a string-theoretical description.  
\par
In spite of these complications, considerable progress has
still been made towards establishing a dS/CFT correspondence
\cite{str2}-\cite{myu}.\footnote{Such studies have, in large part,
 been motivated
by  recent evidence that our universe has a positive cosmological constant 
\cite{per}. Thus, by investigating the dS/CFT duality, one
might hope to understand ``realistic'' quantum gravity  in a holographic way.} 
The crucial points are  as follows:
(i) the CFT is a Euclidean one that lives  on a spacelike boundary at
temporal infinity and (ii) the dS cosmological horizon
adopts  the role played by  the black hole horizon in  the AdS/CFT
 duality.
Some of the more recent studies have considered
 generalizing the Cardy-Verlinde formalism for  a dS (bulk) scenario
\cite{odie2,dan,ogu,cai2,medx,med,hal,myu}. Although
such  treatments have  achieved only qualified success (see discussion below), 
there does indeed appear to be a Cardy-Verlinde-like description
of the boundary entropy in a dS context.  
\par
Of particular interest to  the current paper is  a very recent study
 by Halyo \cite{hal}. (For  earlier,
closely related works, see  Refs.\cite{dan,cai2}.)
 This author 
considered  a Schwarzschild-de Sitter (SdS)  
spacetime  and implemented  a somewhat novel approach in
deriving the   generalized Cardy-Verlinde formula of 
interest.\footnote{The novelty of Ref.\cite{hal}
 being that the derivation does not incorporate the Bekenstein-Hawking entropy
(of the SdS cosmological horizon). 
 However, the derivation does utilize the Hawking temperature,
which implies this  entropy via the first law
of thermodynamics \cite{gh}.  Hence, the novelty is 
really in the presentation, rather than the
derivation {\it per se}.} 
 The premise of this approach  is that
 a Cardy-Verlinde-like   form (for the  entropy of a suitable boundary 
theory)  
can be obtained 
with only three pieces of input:
 the  bulk  metric, the conformal symmetry on the
asymptotic boundary and the Hawking temperature at the cosmological
horizon.
Along with this  derivation, it was shown that   
the boundary theory  can  be interpreted in terms of
Strominger's  realization:   Euclidean  time evolution in a dS bulk   
is dual to a 
renormalization group flow \cite{str}. (Also see Ref.\cite{bbm}.)   
\par
In the current paper, our analysis    begins with a review of Haylo's  
procedure 
in the context of  an SdS-black hole bulk.
We then ``break new ground'' by extending this treatment to a pair of 
interesting cases: the Reissner-Nordstrom-de Sitter (RNdS) black hole
and  a  certain class of asymptotically de Sitter solutions
lacking a  black hole horizon.
This latter class of ``topological'' de Sitter (TdS) spacetimes
was  first considered by Cai, Myung and Zhang \cite{cmz}
and  has since  been  investigated in Refs.\cite{cai2,med,myu}. 
\par 
Early indicators  from the cited studies suggest 
that TdS solutions can be used to circumvent
many of the difficulties  associated with the ``conventional'' 
dS/CFT correspondence.
These problematic points 
 include a negative energy on the boundary (which implies
a non-unitary CFT), a failure to incorporate
the  thermodynamics of the (usual) black hole horizon into the CFT framework,
and formal breakdowns that occur in a dynamical-boundary scenario.
On the other hand, TdS solutions have the undesirable feature
of a naked singularity.\footnote{In fact,
the original motivation for the study of  TdS spacetimes \cite{cmz} was
to test an earlier conjecture on cosmological singularities \cite{bbm}.}
 It is just possible, however, 
that there is some  CFT that  remains well defined while effectively
 describing this singular bulk.
(Noting that a dual   boundary theory is not necessarily 
sensitive to the intricate  details of the corresponding bulk geometry
\cite{bkr}.)
On this basis, we propose that such solutions do indeed
 merit further investigation. 
\par
The rest of the paper is organized as follows. In the next section,
by way of review \cite{hal},
we  derive a generalized form of the Cardy-Verlinde entropy for
a SdS background. Also, a  brief account is
given on  how
 the asymptotic boundary theory
 can be interpreted  in terms of renormalization group flows.
In Section 3,  we generalize the prior formalism to accommodate 
 a RNdS background. That
is, we  consider the implications (at the conformal  boundary)
 when  charge is incorporated into an
asymptotically  dS  background. 
In Section 4, the analytic program is further extended to the case of a TdS 
bulk spacetime.
 We also  elaborate on the distinctions that exist
between these topological solutions and   ``conventional''
asymptotically dS  spacetimes.    
 Section 5 closes with a brief summary and discussion.

\section{Schwarzschild-dS/CFT Correspondence}  

We  begin the formal  analysis
by considering
  a  Euclidean CFT on the asymptotic boundary of a
Schwarzschild-de Sitter (SdS) background.\footnote{Since conformal
symmetry in the bulk is broken by the presence of a black hole,
a prospectively dual boundary theory is, strictly speaking, not necessarily
a conformal one \cite{str}. Nonetheless, for convenience sake,
we will continue to refer to the relevant
 boundary theories as  CFTs.} The  primary
focus  of this section is to derive a generalized form  of the
 Cardy-Verlinde formula
for this scenario. We again note that this is essentially a review
of a derivation found in  Ref.\cite{hal}.
\par
Let us first consider an $n$+2-dimensional SdS metric in
manifestly  static coordinates
\cite{ssv}:
\be
ds^2_{n+2}=-h(r)dt^2+{1\over h(r)}dr^2+r^2d\Omega^2_{n},
\label{1}
\ee
\be
h(r)=1-{r^2\over L^2}-{\omega_{n}M\over r^{n-1}},
\label{2}
\ee
\be
\omega_{n}={16\pi G_{n+2}\over n \V}.
\label{3}
\ee
Here, $L$ is the curvature radius of the  asymptotically dS background
(i.e., $L^{-2}=2\Lambda/n(n+1)$, where $\Lambda$ is
 the positive cosmological constant),
$d\Omega_n^2$
denotes the line element of an $n$-dimensional
 constant-curvature hypersurface
of positive curvature,  
 $\V$ is the volume of this hypersurface, $G_{n+2}$ is
the $n$+2-dimensional Newton constant, and $M$ 
 is a  constant of integration. $M$  can be identified
with the conserved 
 mass of a   SdS black hole  and is always non-negative.
\par
Note that a positive mass and a positively curved
 hypersurface are necessary  conditions for
the existence of an apparent black hole horizon.
In fact, the  allowed range of mass values is given by $0\leq M\leq M_N$,
where:
\be
M_N= {2l^{n-1}\over (n+1)\omega_n}\left({n-1\over n+1}\right)^{{n-1\over 2}}
\label{nar}
\ee
is the mass of an $n$+2-dimensional  Nariai  black hole  \cite{nar}. 
In this Nariai 
case, the black hole horizon coincides  with  the cosmological
horizon, whereas the cosmological horizon typically lies
on the outside (i.e., further from the singularity at $r=0$).
On the other hand, $M=0$  describes a ``pure'' dS space having no
black hole horizon; only  a cosmological horizon at $r=L$.
\par
In general, the cosmological horizon, $r= R\le L$, corresponds 
to  the largest  root
of $h(r)$. This identification  leads to a  defining relation for
the black hole mass:
\be
M={R^{n-1}\over\omega_{n}}-{R^{n+1}\over\omega_{n}L^2}.
\label{mas}
\ee
\par
From a holographic perspective,
the quantity of immediate interest, rather than the mass
of the black hole, is the corresponding excitation of
gravitational energy as measured at ${\cal I}_+$ (i.e.,  future spacelike 
infinity).\footnote{Asymptotically dS spacetimes actually have
a pair of spacelike  asymptotic  boundaries: ${\cal I}_{+}$  
at future infinity  and
 ${\cal I}_{-}$ at past infinity 
\cite{ssv}. 
We will typically be considering the CFT at
future infinity.}
As discussed in Section 1, there are inherent 
 difficulties in defining conserved charges
in an asymptotically dS spacetime.  Nevertheless, 
Balasubramanian {\it et al.} \cite{bbm}
have recently calculated the gravitational energy of  interest
by suitably  adapting the  Brown-York tensor  \cite{by}.
\par
Applying the results of Ref.\cite{bbm} and  subtracting off 
the vacuum-energy contribution 
from
the anomalous Casimir effect (which is non-zero when
$n$+2 is odd), we  obtain the following excitation energy:
\be
E_{SdS}=-M.
\label{ene}
\ee
Significantly,  the negative sign  implies that the presence of
a black hole actually  lowers the total bulk energy  with respect
to the total energy for  a  pure dS spacetime (for which
 the
gravitational excitation has been calibrated to zero).
\par
Let us now consider the  asymptotic form of the 
SdS metric at future spacelike infinity. First, though, we note the following.
For a given choice of coordinates covering  an asymptotically dS
spacetime,  ${\cal I}_+$
can be  defined by the limit $\tau\rightarrow +\infty$;  where
 $\tau$ is an appropriately defined timelike coordinate \cite{ssv}. In
the case of our  static coordinate system (\ref{1}), $r$ becomes
the relevant timelike coordinate (close to this asymptotic boundary), 
as  $t$ and
$r$ switch their usual roles  when an observer
 passes through  the cosmological
horizon.\footnote{Since the  exterior region (with respect to
 the cosmological horizon) is inaccessible to an internally located 
 observer,
we are referring to a hypothetical,  omnipotent  spectator.}
  Hence, the SdS metric (\ref{1}) at ${\cal I}_+$ should be determined
by  way of the following   limit:
\be
\lim_{r\rightarrow\infty}\left[{L^2\over r^2} ds^2_{n+2}\right]
=dt^2+L^2d\Omega^2_{n},
\label{asy}
\ee
\par
Evidently, this asymptotic limit yields a  spacelike metric
describing  a relatively simple  geometry: ${\cal R}\times S^{n}$.
 Moreover, the resulting metric can also be identified with that 
of  an $n$+1-dimensional
 Euclidean CFT (on ${\cal I}_+$).
 It is appropriate, however,  to rescale the boundary
coordinates so that  the radius of the spatial surfaces
coincides with the radial distance of the cosmological horizon \cite{ver}.
(In doing so, we are exploiting the  conformal
symmetry of the asymptotic boundary \cite{mal,gub,wit}.) That
is, $t\rightarrow tL/R$. As a consequence of this rescaling,
it follows that the CFT energy should be ``red shifted''
from $E_{SdS}$
 by the same factor of $L/R$.
\par
On the basis of the above arguments and Eqs.(\ref{mas},\ref{ene}), 
the CFT energy
is given as follows:
\be
E_{CFT}= {nC\over 24}{R^n\over L^{n+1}}\left[1-{L^2\over R^2}\right],
\label{cft}
\ee
where  we have defined:
\be
 C\equiv {3L^n\V \over2\pi G_{n+2}}.
\label{ccc}
\ee 
 $C$ can  readily be
identified with  the Cardy-like ``central charge'' \cite{car}
of the CFT corresponding to
a pure  dS spacetime \cite{str2}. 
We will follow  Ref.\cite{hal}
and presume that $C$  remains   a fixed quantity  for  an  asymptotically
dS theory.
\par
As it turns out,   the CFT energy  can be expressed in terms
of a pair of separable  contributions:
 $E_{CFT}=E_E+E_C$. 
 The first term, 
$E_E= nCR^n/ 24 L^{n+1}$, is
strictly positive and can be identified
as  the extensive energy of an  $n$-dimensional CFT gas \cite{ver}.
It follows that the second   term, $E_C= - E_E L^2/R^2$,
must be a sub-extensive contribution.  Notably, $E_C$
 is always negative and  
can be identified
with  the   Casimir energy of the CFT  \cite{ver}.
Keep in mind  that $E_{CFT}$ must be negative for a 
non-vanishing black hole mass,
since $L> R$ if $M>0$.
\par
Next, let us consider the Hawking temperature associated
with the cosmological horizon.  Applying the well-known procedure
of  Gibbons and Hawking \cite{gh,gh2} along with  Eqs.(\ref{2},\ref{mas}),
 we have:
\be
T_{SdS}=-{1\over 4\pi} \left.{dh\over dr}\right|_{r=R}={R\over 4\pi L^2}
\left[(n+1)-(n-1){L^2\over R^2}\right].
\label{tem}
\ee
It follows that the temperature of the CFT is  given as the red-shifted value
of the above. That is:
\be
T_{CFT}={L\over R}T_{SdS}=
{1\over 4\pi L}
\left[(n+1)-(n-1){L^2\over R^2}\right].
\label{tem2}
\ee
\par
We will  now use a standard thermodynamic relation:
\be
T_{CFT}=\left({\partial E_{CFT}\over \partial S_{CFT}}\right)_{V}
\label{the}
\ee
to obtain the  CFT entropy. More precisely,  we  begin by
 fixing the boundary volume $V=\V R^{n}$ (i.e., fixing  $R$), while
treating $L$ (but not $C$) as a variable quantity. We then obtain
an expression for $\delta S_{CFT}$ by dividing
$\delta E_{CFT}$ with $T_{CFT}$.  As it so happens, the resulting
 variation in entropy
can  be trivially integrated to yield $S_{CFT}$. Note
that there is an arbitrary constant of integration,
which  will always be fixed to  vanish. This choice can be justified
with an appeal to the black hole area law (see below).
\par
Applying the above methodology and
 Eqs.(\ref{cft},\ref{tem2}), we find that
Eq.(\ref{the}) is satisfied
by the following:
\be
S_{CFT}={\pi CR^n\over 6 L^n}={4\pi\over n}R\sqrt{E_E|E_C|}.
\label{ent}
\ee
Remarkably, this CFT entropy  agrees
with the  Bekenstein-Hawking area law \cite{bek,haw}, since
it can easily be verified that
 $S_{CFT}= V /4 G_{n+2}$. (Note that $V$ is clearly the $n$+2-dimensional
generalization of a horizon surface area.)  
However, this should not be interpreted as a derivation of the
Bekenstein-Hawking formula, but rather as a check on consistency
and a demonstration that the CFT entropy is unambiguously defined.
\par
The resulting  CFT entropy can also be identified
as
a modified version of the Cardy formula
 \cite{car}.  
In fact, the only explicit  difference
between Eq.(\ref{ent}) (for SdS)
and the Cardy-Verlinde formula (for Schwarzschild-AdS) \cite{ver}
is the need for absolute value bars.
Note that the above formalism  also holds for a pure dS space.
In this case, one sets $R=L$ to obtain $E_{CFT}=0$, $E_C=-E_E$,
$T_{CFT}=1/2\pi L$ and $S_{CFT}= \V L^n /4 G_{n+2}$.   
\par
It is useful to rewrite this generalized Cardy-Verlinde expression
(\ref{ent}) so that it
more closely resembles the original Cardy formulation.
Let us first define the following:
\be
c\equiv{24\over n}R|E_C|,
\label{def1}
\ee
\be
L_{o}\equiv{1\over n}R E_{CFT}.
\label{def2}
\ee
The modified Cardy-Verlinde formula (\ref{ent})
then takes on the form:
\be
S_{CFT}=2\pi\sqrt{{c\over 6}\left[L_o+{c\over 24}\right]}.
\label{cv}
\ee
As for the dual  CFT of an (asymptotically)  AdS spacetime, $L_o$ 
represents the product of the total energy
and the radius, while  $c/ 24$ is a shift caused by the Casimir effect
\cite{car,ver}. What is of interest  here  are the relative signs: $L_{o}$ is
now a negative quantity (or vanishing  for pure dS),
while the  ``Casimir shift'' 
 is now positive (since
$c\ge 0$).  Conversely,   $L_{o}$ is  positive and the Casimir shift
is negative for 
 the  Schwarzschild-AdS case.
 Note that $c$ is analogous to the Cardy central
charge and, as anticipated, goes to $C$  as $R\rightarrow L$ 
(i.e., for pure dS).
\par
We have observed that, from a  holographic perspective, 
 the  SdS ``picture'' is   quite similar
to that of its Schwarzschild-AdS counterpart. However, there
are some notable  issues that are specific to  the  SdS scenario.
For instance,   $E_{CFT}<0$  
 is a 
precarious outcome,
as it implies that the boundary theory fails to be unitary.
(Note that this negative  energy
is especially problematic   in a  dynamical boundary scenario
\cite{ogu,medx,myu}.)
 Given this 
non-unitarity and  that 
 the SdS spacetime lacks conformal
symmetry  (it is broken by the black hole), there
is no reason to expect, {\it a priori}, that the bulk theory
can be described by a  dual CFT. 
 Furthermore, the boundary entropy, $S_{CFT}$, is bounded
from above by its value for pure dS space (given that
a negative $L_o$ is vanishing in this limit). Such an upper limit
on the accessible degrees of freedom 
 would appear to be  a counter-intuitive result.\footnote{However, we note
 that, from a bulk perspective, such  an upper bound on the entropy is
a well-accepted  quirk of asymptotically dS spacetimes \cite{bou}.}
Finally,  the modified Cardy-Verlinde formula
 fails to incorporate the thermodynamics
of the black hole;  only  the cosmological horizon has been considered.
\par
Some of these complications are effectively negated
when viewed from a perspective that
has been argued for by
 Halyo 
 \cite{hal}.
The premise of  these arguments is based, in large part, on
Strominger's identification \cite{str}  of a dS/CFT-inspired  duality between
Euclidean time evolution  and 
 an appropriate  renormalization group (RG) 
flow.\footnote{Also see Ref.\cite{bbm}. 
For earlier work on RG flows in a holographic setting,
consult, for instance, Refs.\cite{bkr,rg1,rg2}.} 
We briefly summarize this point of view  in the following discussion.
\par
Let us first point out that the SdS black hole is known to be
unstable; this is by virtue of the Hawking temperature
being higher at   the black
hole horizon  than 
at  the cosmological horizon \cite{gh}.
Given this instability, it follows
that the black hole mass, $M$, gradually decreases as time increases.
(Accordingly, the negative  boundary energy, $E_{CFT}$, gradually
increases towards zero.)
After the black hole completely evaporates,  what remains behind  is a
pure dS spacetime, which is,  of course,  perfectly stable.
\par
Let us now consider this evaporation process in 
the  context of a RG flow. Strominger argues
\cite{str} that the time evolution (in the dS bulk) is expected
to be dual with the reverse of a RG flow  on the boundary.\footnote{In
this context, reverse implies that degrees of freedom are being
integrated out as time devolves.}
 That is, a flow
 from an  infrared fixed point (a conformally invariant point with
a relatively  low number
of degrees of freedom) to an ultraviolet fixed point (comparatively  high
number of degrees of freedom).
We can identify the CFT for
 pure dS as an ultraviolet fixed point, 
 given  the obvious conformal symmetry and 
 that it  corresponds to an upper bound on the  entropy. It then follows
that the boundary theory
 for a  SdS black hole (a point of relatively low entropy)
should   evolve with time towards this pure-dS fixed point of
maximal entropy.
Notably, this viewpoint is in perfect agreement  with the thermodynamic
behavior that we discussed above.
\par
 Moreover, since the  boundary
theory must inevitably reach the ultraviolet fixed point,
it can always  be interpreted as an asymptotically  conformal theory.
Of further note,
 as time evolves
(i.e., RG flows from infrared to ultraviolet) the  energy
on the  boundary monotonically increases from a negative value towards
zero. That is,  the boundary theory effectively behaves in 
a  quasi-unitary manner. 
\par
Also of interest, the Nariai black hole assumes
 the role of the (unstable) infrared fixed point in this RG-flow picture.
This follows from the Nariai black hole having the largest
allowable   mass (cf. Eq.(\ref{nar})) 
and, hence,   the smallest (or most negative) allowed values
 for the CFT entropy
and energy. 
\par
Finally, we point out that
one issue   remains conspicuously unresolved; namely, 
the failure of the generalized Cardy-Verlinde formula
to account for the  thermodynamics of the black hole horizon. 
We  will  have more to say on this matter in Section 4.

\section{Reissner-Nordstrom-dS/CFT Correspondence}

In this section, we will apply the prior techniques to
the case of a Reissner-Nordstrom-de Sitter (RNdS)
background.   Such a spacetime represents
the black hole 
 solutions of an  Einstein-Maxwell action  having
 a positive cosmological constant.
\par
We begin here with the RNdS metric for an $n$+2-dimensional spacetime
 in static coordinates
\cite{mann}:
\be
ds^2_{n+2}=-u(r)dt^2+{1\over u(r)}dr^2+r^2d\Omega^2_{n},
\label{x1}
\ee
\be
u(r)=1-{r^2\over L^2}-{\omega_{n}M\over r^{n-1}}+{Q^2\over
r^{2n-2}},
\label{x2}
\ee
where  $Q$  describes
the conserved charge of the associated  black hole and all other parameters
are as previously defined.
\par
Here, there are two possibilities. Either the RNdS black
holes are magnetically charged or electrically charged \cite{bou2}.
Since the  electrically charged black holes can lose their charge
through the emission of  particles (and  thereby decay
into SdS black holes), we will
focus on the former case; thus implying that  the charge
is a fixed quantity.
\par
Typically for $n\geq 2$, there will be   three  positive
(real) roots of $u(r)$; with the  outermost root describing a cosmological
horizon, and the  remaining pair describing  inner and
 outer black hole horizons.
\par 
The   allowed range of mass values (assuming that naked 
singularities are forbidden)
 can now be expressed as $M_{min}\leq M\leq M_{max}$.  Here,
$M_{max}$ corresponds to the so-called charged Nariai solution,
in which case the outermost black hole horizon
coincides with the cosmological horizon. Meanwhile, $M_{min}$
corresponds to an extremal black hole  (i.e., the inner and outer
black hole horizons coincide). In general, these bounds are difficult
to solve analytically.  However,  it is only necessary
to know that they exist and are well defined. Keep in mind, though, that
$M_{min}>0$  and $M_{max}>M_N$ if $Q^2>0$.
\par
Similar to before, we can determine  the location of 
the cosmological horizon, $r=R$,  
 by solving  for the largest  root
of $u(r)$. This identification  yields the following  defining relation for
the black hole mass:
\be
M={R^{n-1}\over\omega_{n}}-{R^{n+1}\over\omega_{n}L^2}+{Q^2
\over\omega_{n}R^{n-1}}.
\label{xmas}
\ee
\par
As discussed in the prior section, one   obtains the CFT energy by taking the
negative   of the red-shifted  mass. That is, $E_{CFT}=-ML/R$ or:
\be
E_{CFT}= {nC\over 24}{R^n\over L^{n+1}}\left[1-{L^2\over R^2}
+{L^2 Q^2\over R^{2n}}\right].
\label{xcft}
\ee
Note that we have also reversed the relative sign on the charge term;
this step follows from 
 the CFT at ${\cal I}_+$ being a Euclidean one \cite{str2}
(which necessitates a complexification of charge along with time \cite{gh2}).
It is convenient to separate
the CFT energy    into three distinct
portions: $E_{CFT}=E_E+E_C+E_{Q}$. The positive
extensive contribution ($E_E$) and the  negative Casimir (or
sub-extensive contribution, $E_C$) are defined exactly
as in the uncharged case.  However, there is
now an additional ``electrostatic'' contribution ($E_Q$),  which
depends on the charge and is clearly non-negative.
This charge-induced portion,  $E_Q$,
 can possibly  be interpreted as a zero-temperature
background energy;  which  implies that it should  make no
contribution to the CFT entropy \cite{cai}.  We will  
demonstrate below that this is
indeed the case.
\par
As for the prior model, the CFT temperature
 corresponds  to the  red-shifted value
of the Hawking temperature at the  cosmological horizon.
So it follows that:
\bea
T_{CFT}&=&-{L\over 4\pi R}\left.{du\over dr}\right|_{r=R}
\nonumber \\ &=&
{1\over 4\pi L}
\left[(n+1)-(n-1){L^2\over R^2}+ (n-1){L^2 Q^2\over R^{2n}}\right].
\label{xtem}
\eea
\par
By applying the
 appropriate thermodynamic relation: 
\be
T_{CFT}=\left({\partial E_{CFT}\over \partial S_{CFT}}\right)_{V,Q}
\label{xthe}
\ee
along with  Eqs.(\ref{xcft},\ref{xtem}),
we should be able to deduce  the form of  the CFT entropy.
Utilizing  the same approach as in the prior section (see the discussion
 leading up to Eq.(\ref{ent})), but now fixing both $V$ and $Q$,
we  ultimately find the following:
\be
S_{CFT}= {\pi CR^n\over 6L^n}={4\pi\over n}R\sqrt{E_E|E_C|}.
\label{xent}
\ee
\par
This CFT  entropy  is formally  identical to  
 that of  the uncharged scenario and is, of course,
 in agreement with the
  Bekenstein-Hawking area law:  $S_{CFT}=\V R^n /4 G_{n+2}$.
As anticipated, the CFT entropy has no explicit knowledge about 
 the electrostatic contribution.\footnote{This entropy does, however,
have an  implicit dependence on the charge. This occurs via
the location of the cosmological horizon; cf. Eq.(\ref{xmas}).}
  This verifies the conjectured status
of $E_Q$ as a zero-temperature background energy (as opposed to
a thermodynamic excitation). That is,
the electrostatic contribution  should
be subtracted from the total energy before  the degrees of
freedom are  totaled up.
\par
As an aside,
 one can define a chemical potential in
the usual way: $\Phi_{CFT}\equiv\partial E_{CFT}/\partial Q$
(with all other parameters being fixed).  This leads to
the  first law of thermodynamics for a constant-volume system:
\be
dE_{CFT}=T_{CFT}dS_{CFT}+\Phi_{CFT}dQ,
\label{fir}
\ee
where $\Phi_{CFT}=nCQ/12L^{n-1}R^{n}$.
\par
Finally, we point out that, because of the duplicity of
the entropy expression, the arguments regarding RG flows
(at the end of Section 2) will  retain their validity in the case
of this charged model.
The only significant difference is the identity of  the fixed points. For
the current analysis, the ultraviolet fixed point
corresponds to the (stable) extremal RNdS black hole 
 and the infrared fixed point is described by the 
(unstable) charged Nariai black hole.  Recall that these are the
solutions of minimal and maximal  allowed mass, respectively.

\section{Topological-dS/CFT Correspondence}  

In this section, we consider a certain brand of asymptotically
dS solutions that were first proposed by Cai, Myung and Zhang
\cite{cmz} and  received further consideration
in Refs.\cite{cai2,med,myu}. These topological-de Sitter (TdS) solutions
can be obtained with a sign reversal   of the mass term in the 
SdS metric. If the mass is kept positive,  this reversal 
 eliminates the black hole horizon and, hence, gives
rise to an undesirable naked singularity. However,   it is within the realm
of possibility that a suitably dual CFT    remains well defined
even  in  the presence of the  bulk singularity. In this regard,
we point out that such a  boundary theory  is not
necessarily sensitive to  the fine-grained details of  
its corresponding  background. (See, for instance, Ref.\cite{bkr}.)
\par
 Although the above argument is  rather speculative,
TdS solutions provide a natural
 means for  resolving
some   of the difficulties associated with the  dS/CFT correspondence.
For instance,  the issue of
 incorporating the thermodynamics
of a dS-black hole horizon (into a  CFT framework)
would   trivially be  negated.  Moreover, the ``mass reversal''
has been shown to induce a positive CFT energy on the asymptotic boundary.  
In view of these desirable features, we argue that such solutions
deserve further investigation and proceed on this basis. 
\par
Let us now  apply  the prior treatment 
to the case of a TdS background geometry.  
The TdS  metric, for an $n$+2-dimensional spacetime, can be
expressed in the following static form
\cite{cmz}:
\be
ds^2_{n+2}=-f(r)dt^2+{1\over f(r)}dr^2+r^2d\Omega^2_{n},
\label{y1}
\ee
\be
f(r)=k-{r^2\over L^2}+{\omega_{n}M\over r^{n-1}}.
\label{y2}
\ee
Here, $M$ is  a mass parameter that is to be regarded as
a  non-negative quantity (in spite
of the sign reversal from Eq.(\ref{2})),
$k$ is a constant of integration that describes the geometry
of the cosmological
 horizon, and all other parameters are as previously defined.
 Without loss of generality, $k$ can be set to +1,0 or -1;
describing a spatial slicing that is respectively elliptic,
flat and hyperbolic.\footnote{By contrast, in the SdS model, the presumed
 existence
of a black hole horizon necessitates the choice of $k=+1$.}
It is interesting to note that the mass parameter  can 
now increase without bound. 
\par
There is   only one positive (real) root of $f(r)$, and
this  locates the position of
the  cosmological horizon, $R$.
On this basis,  the  following relation for
the  mass parameter can be obtained:
\be
M={R^{n+1}\over\omega_{n}L^2}-k{R^{n-1}\over\omega_{n}}.
\label{ymas}
\ee
\par
As usual, we are particularly interested in the gravitational
excitation energy
as measured at ${\cal I}_+$.  By applying the formalism of Refs.\cite{bbm},
Cai {\it et al}. \cite{cmz} obtained the following intuitive result
for this energy:
\be
E_{TdS}= M.
\label{yene}
\ee
Note that the
 anomalous Casimir contribution has, as usual, been subtracted off.
Significantly,  a non-vanishing mass 
now induces an excitation of positive gravitational
energy.  
\par
As priorly discussed, we obtain the  CFT energy  by red shifting
  the gravitational  excitation with a factor of  $L/R$.
That is: 
\be
E_{CFT}= {nC\over 24}{R^n\over L^{n+1}}\left[1-k{L^2\over R^2}\right].
\label{ycft}
\ee
Note that the initial conditions require this energy to be a  non-negative
quantity.
\par
Once again, we are able to  express  this   CFT (total) energy 
 as a sum  of identifiable contributions:
 $E_{CFT}=E_E+E_C$.  
 The extensive  term, 
$E_E= nCR^n/ 24 L^{n+1}$,  is always positive (as was also found
for the SdS theory).
Meanwhile, the Casimir (sub-extensive) contribution is given by:
\be 
 E_C= -k {L^2\over R^2} E_E 
\label{ycas}
\ee
 and can be positive, negative or vanishing; depending on the
choice of $k$. (By contrast, the SdS Casimir energy is 
always negative.)
\par
As demonstrated before,
the CFT temperature is  obtainable as the red-shifted
value of the Hawking temperature at the cosmological horizon.
More explicitly:
\be
T_{CFT}=-{L\over 4\pi R}\left.{df\over dr}\right|_{r=R}
={1\over 4\pi L}
\left[(n+1)-(n-1)k{L^2\over R^2}\right].
\label{ytem2}
\ee
\par
For the purpose of calculating the CFT entropy,
let us now reconsider the following relation:
\be
T_{CFT}=\left({\partial E_{CFT}\over \partial S_{CFT}}\right)_{V}.
\label{ythe}
\ee
Adopting the approach of the prior sections, we obtain:
\be
S_{CFT}={\pi CR^n\over 6 L^n}={4\pi\over n}R
\sqrt{E_E\left|{E_C\over k}\right|}.
\label{yent}
\ee
Note that the quantity $\left|k^{-1}E_C\right|$  translates
to  $nCR^{n-2}/24 L^{n-1}>0$, regardless of the choice
for $k$. That is, the CFT entropy contains no explicit
information about this TdS geometrical parameter. (Although
$k$ does have implicit influence  via $R$.)
\par
As found  for the previous models,
$S_{CFT}$ agrees
with the anticipated Bekenstein-Hawking form:  $V /4 G_{n+2}$.
Moreover, 
the entropy  can be identified
as
a  (slightly) modified version of the Cardy-Verlinde formula
 \cite{car, ver}. 
Note that the above formalism  also holds for  a pure dS space.
In this case, one sets $R=L$ and $k=1$ to obtain $E_{CFT}=0$, $E_C=-E_E$,
$T_{CFT}=1/2\pi L$ and $S_{CFT}= \V L^n /4 G_{n+2}$.   
\par
As  for the SdS model,
it is informative to re-express $S_{CFT}$ 
(\ref{yent}) into a form that resembles
 the original Cardy relation.
With this in mind, let us define:
\be
c\equiv{24\over n}R\left|{E_C\over k}\right|,
\label{ydef1}
\ee
\be
L_{o}\equiv{1\over n}R E_{CFT}.
\label{ydef2}
\ee
The modified Cardy-Verlinde formula (\ref{yent})
then becomes:
\be
S_{CFT}=2\pi\sqrt{{c\over 6}\left[L_o+k{c\over 24}\right]}.
\label{ycv}
\ee
Notably, this relation is closer to the original Cardy-Verlinde
formulation \cite{car,ver} than was  found  for the SdS analysis (\ref{cv}).
First of all, $L_o$ is  a positive quantity for the TdS-bulk  case. 
(Conversely, it is negative for the SdS scenario.)
Secondly, consider the relative shift as induced by the Casimir effect.
Given  a dS context, this can only be negative for a TdS-bulk spacetime.
This occurs when
 $k=-1$, which  happens to be  the condition for
 a positive Casimir energy (cf. Eq.(\ref{ycas})). In this sense,
it is the case of a  hyperbolic horizon geometry which most closely
resembles the  Schwarzschild-AdS ``template''.
\par
Let us now recall the discussion on  RG flows at the end of Section 2.
 For the current TdS scenario,  such an interpretation
is  hindered  by difficulties in  
identifying the relevant fixed points.   One notable exception
being the infrared fixed point for the $k=1$ case, as
the associated pure-dS spacetime (i.e., the $M=0$ solution)
conveniently imposes  a lower bound
on  $S_{CFT}$.  Alternatively, the massless limit is not so well defined
if  $k=0$ or $-1$. For these topologies, the cosmological
horizon disappears when (or before) $M$ goes to zero.
However, the existence of such exotic bulk geometries is not necessarily
an issue, as we argue below.
\par
In interpreting the boundary theory of  a TdS background,  we
first take note of
the key finding:  
 the  CFT (total) energy
 is a strictly non-negative quantity. This implies that    
 the  boundary theory is   
 a unitary one that is capable of describing bulk spacetime  events;
 for instance, the emission of Hawking radiation from the
cosmological horizon. 
\par
 Let us now consider the following picture. In the distant past,
the TdS bulk  is assumed to be in a relatively massive state (i.e., large $M$
and, thus, large $S_{CFT}$). It follows that the cosmological horizon
will necessarily  radiate until  it  achieves a state of
thermal equilibrium with the emitted radiation. (Such an equilibrium
state follows  by virtue of there being no black hole
horizon in this model.) When this occurs, 
there will be zero net radiation and the bulk will have
settled into a state of comparatively low mass (i.e., small $M$
and, thus, small $S_{CFT}$). However, we
conjecture that this ``final'' value of  mass is 
 not so low that the  cosmological horizon (in the case
of $k=0$ or $-1$) will be  jeopardized (as discussed  above).  To put
it another way,
time evolution in the bulk is dual with a RG flow  from
ultraviolet to infrared  boundary points that are
at least effectively fixed.
\par
Along with the apparent unitarity of the boundary theory, the above picture
is supported by  an intuitively satisfying  correspondence:
 the boundary degrees of freedom are directly
correlated with the  gravitational mass in the bulk.
 This  is at least the case
 for $k=-1$ and $k=0$, where 
it is clear that
  an increasing (decreasing) $M$ 
always  corresponds to  an 
increasing
(decreasing) $S_{CFT}$.
\par
On the other hand, this  monotonic mass-entropy   
trend is not so evident  for the case
of
 $k=+1$.  With this topological  choice, a  
numerical analysis will  likely be required
to establish the true mass-entropy relationship. However, the negative
Casimir energy for $k=+1$  suggests that this  may  be
the ``least physical'' case of the three.  Interestingly, 
peculiar behavior linked to
a negative Casimir energy 
 has been  detected  for another background:  
 a ``topological'' AdS black hole with $k=-1$ \cite{cai}.\footnote{From
 an AdS perspective, 
topological  refers to asymptotically AdS-black hole solutions 
 having  either a hyperbolic ($k=-1$) or flat ($k=0$)  hypersurface
\cite{bir}. Note that
 $k=+1$ describes the ``usual'' Schwarzschild-AdS black hole.} 
\par
 In some sense,
TdS solutions are  ``mirror images'' of their AdS analogues, with
 the cosmological horizon replacing the AdS black hole horizon.
Furthermore, in view of the relative sign of the Casimir energy, 
a TdS hyperbolic horizon  ($k=-1$) should be
 regarded as the ``reflection'' of an AdS spherical horizon  ($k=+1$)
and  {\it vice versa}.

\section{Conclusion}

In summary, we have considered  generalized forms of the Cardy-Verlinde 
entropy formula \cite{car,ver} in the context of asymptotically de Sitter 
spacetimes.  Towards this end,  we have applied a program of study that
 is based on
a prior work by Halyo \cite{hal}. The premise of this  methodology
is that one can derive the  entropy of an appropriately dual CFT 
by using only the bulk metric, the asymptotic conformal
symmetry and the  Hawking radiation of the cosmological
horizon \cite{gh}.  The validity of this
approach was substantiated by the reproduction (in all examined
cases) of
the Bekenstein-Hawking area law, which is
 expected to be a fundamental feature of any quantum theory of gravity
\cite{bek,haw}.  
\par
Our analysis considered three distinct cases
having dS asymptotics. First of all, we  examined the Schwarzschild-de Sitter
black hole, which was essentially  a review of the  originating
work \cite{hal}.  In particular, we derived the  entropy of 
the dual CFT, 
and  demonstrated
that it adopts a Cardy-Verlinde-like form.  We also  argued that, 
 in spite of apparent conceptual difficulties (such as a non-unitary
CFT), the proposed  SdS/CFT duality can  still   fit into
a  self-consistent framework.
These arguments  were based on an observation by Strominger \cite{str}:
Euclidean time evolution in a de Sitter space is dual
to a renormalization group  flow from an infrared to an ultraviolet
fixed point. 
\par
Secondly, we considered the Reissner-Nordstrom-de Sitter black hole.
We found that the addition of charge into the model
results in an ``electrostatic''  contribution to
the CFT energy. However,  the charge made no explicit
contribution to the boundary entropy, which implies that the electrostatic
energy can be interpreted  as a zero-temperature background 
 (rather than a thermodynamic excitation).
This outcome  was  anticipated \cite{cai}, but not necessarily obvious.
\par
Finally, we  applied the analytic  program
to a special class of ``topological'' de Sitter solutions.
Such solutions correspond to a sign  reversal in the mass
term of the corresponding Schwarzschild-dS scenario.
For this case, the  CFT entropy  most closely
resembled the original  Cardy-Verlinde formulation
(for AdS spacetimes) \cite{car,ver}. Furthermore,
the  boundary energy was found to be positive, in stark contrast
to the priorly studied cases. This is a desirable outcome,
as it suggests a unitary  theory at the conformal boundary.
With this property of unitarity, we were able to  conceptualize
a duality between an apparent flow in the boundary theory
 and time  evolution in the TdS bulk.      
\par
Let us further point out two  notable failures of  SdS-type models
(with regard to the proposed holographic  duality)
that TdS solutions can seemingly resolve. These are as follows:
 (i) the issue of how to incorporate
the properties of the black hole horizon into the CFT  thermodynamic
 relations (there is no such horizon for TdS spacetimes)
and (ii) a negative CFT energy can become especially problematic
 in the context of a dynamic-boundary scenario \cite{medx,med}.\footnote{We
note, however, that Myung has recently argued  against
both  SdS and TdS backgrounds 
in such  a cosmological setting \cite{myu}.}
\par
In view of the above considerations,  TdS spacetimes appear to
be promising candidates for the realization of a dS/CFT correspondence. 
However,
we again stress  that TdS solutions have a naked singularity
as an inevitable consequence.  It is possible that 
a  well-defined dual CFT  exists  in spite of this singular behavior; 
however,  until this existence can be established, the outcomes of
Section 4  should  be regarded
as speculative.
\par
In conclusion,
 a holographic interpretation of
 asymptotically de Sitter spacetimes 
continues to have a few unresolved issues; whether it is viewed from 
the perspective of a  black hole (SdS, RNdS) or a topological (TdS) bulk.
 Nonetheless, in weighing all the evidence,
we find that the results of this analysis 
 comes out in support
of the proposed dS/CFT correspondence.

\section{Acknowledgments}
\par
The author  would like to thank  V.P.  Frolov  for helpful
conversations.

\end{document}